# One-Shot Achievable Secrecy Rate Regions for Quantum Interference Wiretap Channel

Hadi Aghaee
Faculty of Electrical Engineering
K. N. Toosi University of Technology
Tehran, Iran
Email: Aghaee_Hadi@email.kntu.ac.ir

Bahareh Akhbari
Faculty of Electrical Engineering
K. N. Toosi University of Technology
Tehran, Iran
Email: akhbari@eetd.kntu.ac.ir

*Abstract*—In this paper, we want to derive achievable secrecy rate regions for quantum interference channel with classical inputs under one-shot setting. The main idea to this end is to use the combination of superposition and rate splitting for encoding scheme and constructing a decoding scheme based on simultaneous decoding.

I. INTRODUCTION

The physical layer security was introduced by Shannon for the first time [1]. After that, the wiretap channel was presented by Wyner, in which a sender transmits its message to a legitimate receiver in the presence of a passive eavesdropper [2]. Moreover, Csiszár and Körner introduced the broadcast channel with confidential messages [3].

However, the physical layer security problems have been extended to multi-terminal channels like multiple access channels (MACs), Interference channels (ICs), relay channels, etc., due to their importance and their usage in practical systems [4-10].

In recent decades, with development in quantum data processing and its applications, a significant effort has begun to use the natural features of quantum mechanics to improve communication. Some of these features are as follows: entanglement, uncertainty, no-cloning theorem, superposition, etc. [11]. These natural features help the communication to be faster and more secure.

Moreover, the security problem plays a critical role in quantum communication and devotes a considerable part of the research to itself. In this regard, the quantum wiretap channel (QWTC) was firstly introduced in [12] and [13].

Then, secrecy constraints are extended to multi-user quantum channels such as quantum interference channel (QIC) [14] and quantum multiple access channel (QMAC) [15-18]. The interference phenomenon is one of the major problems in communication systems.

In this paper, we derive some achievable secrecy rate regions for quantum interference channel with classical inputs.

One of the major open problems in the quantum information theory is related to the simultaneous decoder for quantum channels with three or more senders (i.e., jointly typical decoder). However, this problem has been solved for some cases, such as the min-entropy case and the case of the quantum multiple access channels (QMACs), in which the output systems are commutative [19]. Therefore, in the independent and identical distributed (i.i.d.) case, we have to use successive decoding combined with time-sharing. In contrast, for the one-shot case, we have to use the simultaneous decoder. Sen proved a joint typicality lemma which is helpful to decode any number of messages simultaneously in the one-shot case [19].

In this paper, we want to study secure communication over a classical-quantum interference wiretap channel (C-QI-WTC) under the one-shot setting. Up to the best knowledge, it is the first time that this channel is studied. Even in the classical case, the security problem of interference channel has been investigated under a different scenario called interference channel with confidential messages. Also, another feature of our problem is that the channel is considered under the one-shot setting. This choice is due to the fact that there is not a proven joint typicality lemma in the asymptotic i.i.d. case for general quantum channels (i.e., quantum channels with any number of senders). Therefore, all of the obtained results are new, and the proposed strategies in the paper can be applied to the classical interference channel.

The paper is organized as follows: In Section II, some seminal definitions are presented. In Section III, the main channel and information processing tasks are presented. In Section IV, the results and main theorems are presented. Section V is dedicated to discussion and future works.

II. PRELIMINARIES

Let A (Alice) and B (Bob) be two quantum systems. These quantum systems can be denoted by their corresponding Hilbert spaces as $\mathcal{H}^A$, $\mathcal{H}^B$. The states of the above quantum systems are presented as density operators $\rho^A$ and $\rho^B$, respectively, while the shared state between Alice and Bob is denoted by $\rho^{AB}$. A density operator is a positive semidefinite operator with a unit trace. Alice or Bob's state can be defined by a partial trace operator over the shared state. The partial trace is used to model the lack of access to a quantum system. Thus, Alice's density operator using partial trace is $\rho^A = Tr_B\{\rho^{AB}\}$, and Bob's density operator is $\rho^B = Tr_A\{\rho^{AB}\}$. We use $|\psi\rangle^A$ to denote the pure state of system A. The corresponding density operator is $\psi^A = |\psi\rangle\langle\psi|^A$. The von Neumann entropy of the state $\rho^A$ is defined by $H(A)_\rho = -Tr\{\rho^A \log \rho^A\}$. For an arbitrarily state such as $\sigma^{AB}$, the quantum conditional entropy is defined by $H(A|B)_\sigma = H(A,B)_\sigma - H(B)_\sigma$. The quantum mutual information is defined by $I(A;B)_\sigma = H(A)_\sigma + H(B)_\sigma - H(A,B)_\sigma$, and the conditional quantum mutual information is defined by:

$$I(A;B|C)_\sigma = H(A|C)_\sigma + H(B|C)_\sigma - H(A,B|C)_\sigma$$

Quantum operations can be denoted by *completely positive trace-preserving* (CPTP) maps $\mathcal{N}^{A\to B}$. The CPTP maps accept input states in A and output states in B. The distance between two quantum states, such as A and B is defined by trace distance. The trace distance between two arbitrarily states such as $\sigma$ and $\rho$ is:

$$\|\sigma - \rho\|_1 = Tr|\sigma - \rho| \quad (1)$$

where $|\Psi| = \sqrt{\Psi^\dagger \Psi}$. This quantity is zero for two similar and perfectly distinguishable states.

*Fidelity* is defined as $F(\rho, \sigma) = \left\|\sqrt{\rho}\sqrt{\sigma}\right\|_1^2$, and *purified distance* is a metric on $\mathcal{D}(\mathcal{H})$ and is defined as $P(\rho, \sigma) := \sqrt{1 - F(\rho, \sigma)^2}$. Most of the above definitions are given from [20].

***Definition 1:*** *(Hypothesis testing mutual information)* Hypothesis testing mutual information is denoted by $I_H^\epsilon(X;Y) := D_H^\epsilon(\rho_{XY} \| \rho_X \otimes \rho_Y), \epsilon \in (0,1)$ and is considered as *quantum hypothesis testing divergence* [21] where $D_H^\epsilon(.\|.)$ is *hypothesis testing relative entropy* [21]. $\rho^{\mathcal{H}_X \mathcal{H}_Y}$ is the joint state of input and output over their Hilbert spaces $(\mathcal{H}_X, \mathcal{H}_Y)$, and it can be shown as $\rho_{XY}$:

$$\rho_{XY} = \sum_x p_X(x)|x\rangle\langle x|_X \otimes \rho_Y^x$$

where $p_X$ is the input distribution.

***Definition 2:*** *(Quantum relative entropy [22]):* Consider states $\rho_X, \sigma_X \in \mathcal{D}(\mathcal{H}_X)$. The Quantum relative entropy is defined as:

$$D(\rho_X\|\sigma_X) := \begin{cases} Tr\{\rho_X[\log_2 \rho_X - \log_2 \sigma_X]\} & supp(\rho_X) \subseteq supp(\sigma_X) \\ +\infty & otherwise \end{cases}$$

where $supp(\sigma_X)$ refers to the *set-theoretic support* of $\sigma$. $supp(\sigma)$ is the subspace of $\mathcal{H}$ spanned by all eigenvectors of $\sigma$ with non-zero eigenvalues.

**Fact**: The following relation exists between the quantum relative entropy and hypothesis testing relative entropy for $\epsilon \in (0,1)$ [21]:

$$D_H^\epsilon(\rho_X\|\sigma_X) \leq \frac{1}{1-\epsilon}[D(\rho_X\|\sigma_X) + h_b(\epsilon)]$$

where $h_b(\epsilon) := -\epsilon \log_2 \epsilon - (1-\epsilon)\log_2(1-\epsilon)$ is a binary entropy function.

***Definition 3:*** *(Max mutual information [23])* Consider a bipartite state $\rho_{XY}$ and a parameter $\epsilon \in (0,1)$. The max mutual information can be defined as follows:

$$I_{max}(X;Y)_\rho := D_{max}(\rho_{XY}\|\rho_X \otimes \rho_Y)_\rho$$

where $\rho$ refers to the state $\rho_{XY}$ and $D_{max}(\|)$ is the *max-relative entropy* [24] for $\rho_X, \sigma_X \in \mathcal{H}_X$:

$$D_{max}(\rho_X\|\sigma_X) := \inf\{\gamma \in \mathbb{R}: \rho_X \leq 2^\gamma \sigma_X\}$$

***Definition 4:*** *(Quantum smooth max relative entropy [24])* Consider states $\rho_X, \sigma_X \in \mathcal{D}(\mathcal{H}_X)$ and $\epsilon \in (0,1)$. The quantum smooth max relative entropy is defined as:

$$D_{max}^\epsilon(\rho_X\|\sigma_X) := \inf_{\rho_X' \in \mathcal{B}^\epsilon(\rho_X)} D_{max}(\rho_X'\|\sigma_X)$$

where $\mathcal{B}^\epsilon(\rho_X) := \{\rho_X' \in \mathcal{D}(\mathcal{H}_X): P(\rho_X', \rho_X) \leq \epsilon\}$ is $\epsilon$-ball for $\rho_{XY}$.

***Definition 5:*** *(Quantum smooth max mutual information [23])* Consider $\rho_{XY} := \sum_{x \in \mathcal{X}} P_X(x)|x\rangle\langle x|_X \otimes \rho_Y^x$ as a classical-quantum state and a parameter $\epsilon \in (0,1)$. The smooth max mutual information between the systems $X$ and $Y$ can be defined as follows:

$$I_{max}^\epsilon(X;Y) := \inf_{\rho_{XY}' \in \mathcal{B}^\epsilon(\rho_{XY})} D_{max}(\rho_{XY}'\|\rho_X \otimes \rho_Y)$$
$$= \inf_{\rho_{XY}' \in \mathcal{B}^\epsilon(\rho_{XY})} I_{max}(X;Y)_{\rho'},$$

where $\mathcal{B}^\epsilon(\rho_{XY}) := \{\rho_{XY}' \in \mathcal{D}(\mathcal{H}_X \otimes \mathcal{H}_Y): P(\rho_{XY}', \rho_{XY}) \leq \epsilon\}$ is $\epsilon$-ball for $\rho_{XY}$.

***Definition 6:*** *(Conditional smooth hypothesis testing mutual information [25])* Consider $\rho_{XYZ} := \sum_{z \in Z} P_Z(z)|z\rangle\langle z|_Z \otimes \rho_{XY}^z$ be a tripartite classical-quantum state and $\epsilon \in (0,1)$. We define,

$$I_H^\epsilon(X;Y|Z)_\rho := \max_{\rho'} \min_{z \in supp(\rho_Z')} I_H^\epsilon(X;Y)_{\rho_{XY}^z},$$

where maximization is over all $\rho_Z' = \sum_{z \in Z} p_Z(z)|z\rangle\langle z|_Z$ satisfying $P(\rho_Z', \rho_Z) \leq \epsilon$.

***Definition 7:*** *(Conditional smooth max mutual information [25])* Consider $\rho_{XYZ} := \sum_{z \in Z} P_Z(z)|z\rangle\langle z|_Z \otimes \rho_{XY}^z$ be a tripartite classical-quantum state and $\epsilon \in (0,1)$. We define,

$$I_{max}^\epsilon(X;Y|Z)_\rho := \max_{\rho'} \min_{z \in supp(\rho_Z')} I_{max}^\epsilon(X;Y)_{\rho_{XY}^z},$$

where maximization is over all $\rho_Z' = \sum_{z \in Z} p_Z(z)|z\rangle\langle z|_Z$ satisfying $P(\rho_Z', \rho_Z) \leq \epsilon$.

***Definition 8:*** *(Quantum Rényi relative entropy of order $\alpha$ [21])* For a state $\rho \in \mathcal{D}(\mathcal{H})$ and a positive semidefinite operator $\sigma$, the *quantum Rényi relative entropy of order $\alpha$*, where $\alpha \in [0,1) \cup (1, +\infty)$ is defined as:

$$D_\alpha(\rho\|\sigma) \equiv \frac{1}{\alpha - 1} \log_2 Tr\{\rho^\alpha \sigma^{1-\alpha}\}$$

Also, *Rényi entropy of order $\alpha$* can be defined as follows:

$$H_\alpha(A)_\rho \equiv \frac{1}{1-\alpha} \log_2 Tr\{\rho_A^\alpha\}$$

***Definition 9:*** *(One-shot inner bound of a classical-quantum multiple access channel)* [19] A two user C-QMAC under the one-shot setting is a triple $(\mathcal{X}_1 \times \mathcal{X}_2, \mathcal{N}_{X_1 X_2 \to Y}(x_1, x_2) \equiv \rho_Y^{x_1 x_2}, \mathcal{H}_Y)$, where $\mathcal{X}_1$ and $\mathcal{X}_2$ are the alphabet sets of two classical inputs, and $Y$ is the output system. $\rho_{x_1 x_2}^Y$ is a quantum state, and the channel has a completely positive trace-preserving map (CPTP) $\mathcal{N}_{X_1 X_2 \to Y}$.

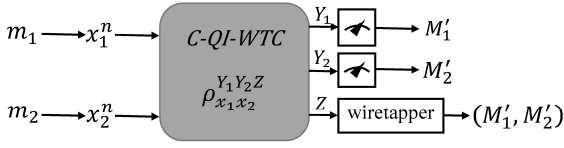

*Figure 1. The C-QI-WTC model*

Considering the joint typicality lemma introduced in [Corollary 4, 19], the one-shot inner bound of a C-QMAC is as follows:

$$R_1 \leq I_H^\epsilon(X_1:X_2Y)_\rho - 2 + \log \epsilon$$

$$R_2 \leq I_H^\epsilon(X_2:X_1Y)_\rho - 2 + \log \epsilon$$

$$R_1 + R_2 \leq I_H^\epsilon(X_1X_2:Y)_\rho - 2 + \log \epsilon$$

where $I_H^\epsilon(.)$ is the hypothesis testing mutual information defined in Definition 1 with respect to the controlling state:

$$\rho_{QX_1X_2Y} := \sum_{qx_1x_2} p(q)p(x_1|q)p(x_2|q)|qx_1x_2\rangle\langle qx_1x_2|^{QX_1X_2} \otimes \rho_Y^{x_1x_2}$$

and $Q$ is a time-sharing variable.

Note that $I_H^\epsilon(:)$ is the difference between a *Rényi entropy* of order two and a conditional quantum entropy.

### III. CHANNEL MODEL

A two-user C-QI-WTC is a triple $(\mathcal{X}_1 \times \mathcal{X}_2, \mathcal{N}^{X_1X_2 \to Y_1Y_2Z}(x_1, x_2) \equiv \rho_{x_1x_2}^{Y_1Y_2Z}, \mathcal{H}^{Y_1} \otimes \mathcal{H}^{Y_2} \otimes \mathcal{H}^Z)$, where $\mathcal{X}_i, i \in \{1,2\}$ denote the input alphabet sets, and $Y_1, Y_2, Z$ denote the output systems ($Y_1, Y_2$ denote the channel outputs at the two legitimate receivers and $Z$ is the channel outputs at the eavesdropper). $\rho_{x_1x_2}^{Y_1Y_2Z}$ is the system output's quantum state. Each user wants to transmit its message as secure as possible over a C-QI-WTC to its intended receiver.

The main channel (i.i.d. case) is illustrated in Figure 1.

Consider the main channel illustrated in Figure 1 under the one-shot setting. Each user chooses its message $m_i; i \in \{1,2\}$ from its message set $\mathcal{M}_i = [1:|\mathcal{M}_i| = 2^{R_i}]; i \in \{1,2\}$, and send it over a C-QI-WTC. The users also use two junk variables $k_i; i \in \{1,2\}$ from two amplification sets $\mathcal{K}_i = [1:|\mathcal{K}_i| = 2^{\hat{R}_i}]; i \in \{1,2\}$ for randomizing Eve's knowledge.

We have two doubly indexed codebooks $x_1(m_1, k_1)$ and $x_2(m_2, k_2)$ for user-1 and user-2, respectively. The above channel can be divided into two sub C-QMA-WTCs (one from both users to $(Y_1, Z)$ and another from both users to $(Y_2, Z)$).

### IV. MAIN RESULTS

In this section, we present the main results.

***Theorem 1:*** *(One-shot achievable rate region for C-QI-WTC)* Consider a two-user C-QI-WTC which accepts $X_1$ and $X_2$ as inputs and $Y_1, Y_2$ and $Z$ as outputs. $\rho_{x_1x_2}^{Y_1Y_2Z}$ is the channel density operator. For any fixed $\epsilon \in (0,1), \epsilon' \in (0, \delta')$ and $\delta, \delta'$ such that $\delta, \delta' > 0$, the rate pair $R_i = \log|\mathcal{M}_i| + \delta, i \in \{1,2\}$ is achievable to satisfy the following inequalities:

$$R_1 \leq \min\{I_H^\epsilon(X_1:X_2Y_1|Q)_\rho, I_H^\epsilon(X_1:X_2Y_2|Q)_\rho\}$$
$$- I_{max}^\eta(X_1:Z|Q)_\rho + \log \epsilon - 1 - \log \frac{3}{\epsilon'^3}$$
$$+ \frac{1}{4} \log \delta$$

$$R_2 \leq \min\{I_H^\epsilon(X_2:X_1Y_1|Q)_\rho, I_H^\epsilon(X_2:X_1Y_2|Q)_\rho\}$$
$$- I_{max}^\eta(X_2:ZX_1|Q)_\rho + \log \epsilon - 1 - \log \frac{3}{\epsilon'^3}$$
$$+ \frac{1}{4} \log \delta$$

$$R_1 + R_2 \leq \min\{I_H^\epsilon(X_1X_2:Y_1|Q)_\rho, I_H^\epsilon(X_1X_2:Y_2|Q)_\rho\}$$
$$- I_{max}^\eta(X_1:Z|Q)_\rho - I_{max}^\eta(X_2:ZX_1|Q)_\rho$$
$$+ \log \epsilon - 1 - 2\log \frac{3}{\epsilon'^3} + \frac{1}{2}\log \delta + \mathcal{O}(1)$$

where $\eta = \delta' - \epsilon'$ and the union is taken over input distribution $p_Q(q)p_{X_1|Q}(x_1|q)p_{X_2|Q}(x_2|q)$. $Q$ is the time-sharing random variable, and all of the mutual information quantities are taken with respect to the following state:

$$\rho^{QX_1X_2Y_1Y_2Z} \equiv$$
$$\sum_{q,x_1,x_2} p_Q p(q) p_{X_1|Q}(x_1|q) p_{X_2|Q}(x_2|q) |q\rangle\langle q|^Q$$
$$\otimes |x_1\rangle\langle x_1|^{X_1} \otimes |x_2\rangle\langle x_2|^{X_2}$$
$$\otimes \rho_{x_1x_2}^{Y_1Y_2Z} \qquad (4)$$

*Proof*: See Appendix A.

*Sketch of proof*: The channel can be split into two sub-QMA-WTCs with classical inputs. One from $(X_1, X_2)$ to $(Y_1, Z)$ and another from $(X_1, X_2)$ to $(Y_2, Z)$. Using the proposed method by El-Gamal and H. Kim [26] helps to prove this theorem.

Theorem 1 gives the simplest achievable rate region for C-QI-WTC under the one-shot setting. Without considering the secrecy constraints, Han and Kobayashi obtained the best achievable rate region for interference channel (i.i.d. setting) using rate splitting that the messages are split into common and personal messages. This technique is extended to the quantum case with some limits [14]. Using the Han-Kobayashi's technique, the message $X_i$ is split into $X_{i0}$ (common part) and $X_{ii}$ (personal part), where $i \in \{1,2\}$.

The structure of the C-QI-WTC under the Han-Kobayashi's setting is illustrated in Figure 2. The following channel can be divided into two separate sub 3-user C-QMA-WTCs: one from $(X_{10}, X_{11}, X_{20})$ to $(Y_1, Z)$ and another from $(X_{20}, X_{22}, X_{10})$ to $(Y_2, Z)$.

As mentioned before, there is not a proven quantum simultaneous decoder for decoding three or more messages in general and it remains a conjecture (except some cases such as the commutative version of output states and min-entropy cases [14]).

$$\{I_{max}^{\eta}(m_{10}m_{11}m_{20}m_{22}:Z)_{\rho} \leq \varepsilon_3 | I_{max}^{\eta}(m_{10}m_{11}m_{20}:Z)_{\rho} \leq \varepsilon_1, I_{max}^{\eta}(m_{10}m_{20}m_{22}:Z)_{\rho} \leq \varepsilon_2\} \quad (5)$$

where $\varepsilon_1, \varepsilon_2$ and $\varepsilon_3$ are arbitrary small numbers.

$$R_1 \leq I_H^{\epsilon}(X_{10}X_{11}:Y_1X_{20})_{\rho} - I_{max}^{\delta'-\epsilon'}(X_{10}:Z)_{\rho} - I_{max}^{\delta'-\epsilon'}(X_{11}:ZX_{10}X_{20})_{\rho} - 2\log\frac{3}{\epsilon'^3} + \frac{1}{2}\log\delta' + \log\epsilon - 2 + \mathcal{O}(1) \quad (6)$$

$$R_1 \leq I_H^{\epsilon}(X_{11}:Y_1X_{10}X_{20})_{\rho} + I_H^{\epsilon}(X_{10}:Y_2X_{20}X_{22})_{\rho} - I_{max}^{\delta'-\epsilon'}(X_{10}:Z)_{\rho} - I_{max}^{\delta'-\epsilon'}(X_{11}:ZX_{10}X_{20})_{\rho} - 2\log\frac{3}{\epsilon'^3} + \frac{1}{2}\log\delta'$$
$$+ 2\log\epsilon - 4 + \mathcal{O}(1) \quad (7)$$

$$R_2 \leq I_H^{\epsilon}(X_{20}X_{22}:Y_2X_{10})_{\rho} - I_{max}^{\delta'-\epsilon'}(X_{20}:ZX_{10})_{\rho} - I_{max}^{\delta'-\epsilon'}(X_{22}:ZX_{10}X_{11}X_{20})_{\rho} - 2\log\frac{3}{\epsilon'^3} + \frac{1}{2}\log\delta' + \log\epsilon - 2$$
$$+ \mathcal{O}(1) \quad (8)$$

$$R_2 \leq I_H^{\epsilon}(X_{20}:Y_1X_{10}X_{11})_{\rho} + I_H^{\epsilon}(X_{22}:Y_2X_{10}X_{20})_{\rho} - I_{max}^{\delta'-\epsilon'}(X_{20}:ZX_{10})_{\rho} - I_{max}^{\delta'-\epsilon'}(X_{22}:ZX_{10}X_{11}X_{20})_{\rho} - 2\log\frac{3}{\epsilon'^3}$$
$$+ \frac{1}{2}\log\delta' + 2\log\epsilon - 4 + \mathcal{O}(1) \quad (9)$$

$$R_1 + R_2 \leq I_H^{\epsilon}(X_{11}:Y_2X_{10}X_{20})_{\rho} + I_H^{\epsilon}(X_{10}X_{11}X_{20}:Y_2)_{\rho} - I_{max}^{\delta'-\epsilon'}(X_{10}:Z)_{\rho} - I_{max}^{\delta'-\epsilon'}(X_{11}:ZX_{10}X_{20})_{\rho}$$
$$- I_{max}^{\delta'-\epsilon'}(X_{20}:ZX_{10})_{\rho} - I_{max}^{\delta'-\epsilon'}(X_{22}:ZX_{10}X_{11}X_{20})_{\rho} - 4\log\frac{3}{\epsilon'^3} + \log\delta' + 2\log\epsilon - 4 + \mathcal{O}(1) \quad (10)$$

$$R_1 + R_2 \leq I_H^{\epsilon}(X_{11}:Y_1X_{20}X_{10})_{\rho} + I_H^{\epsilon}(X_{22}X_{20}X_{10}:Y_2)_{\rho} - I_{max}^{\delta'-\epsilon'}(X_{10}:Z)_{\rho} - I_{max}^{\delta'-\epsilon'}(X_{11}:ZX_{10}X_{20})_{\rho}$$
$$- I_{max}^{\delta'-\epsilon'}(X_{20}:ZX_{10})_{\rho} - I_{max}^{\delta'-\epsilon'}(X_{22}:ZX_{10}X_{11}X_{20})_{\rho} - 4\log\frac{3}{\epsilon'^3} + \log\delta' + 2\log\epsilon - 4 + \mathcal{O}(1) \quad (11)$$

$$R_1 + R_2 \leq I_H^{\epsilon}(X_{11}X_{20}:Y_1X_{10})_{\rho} + I_H^{\epsilon}(X_{22}X_{10}:Y_2X_{20})_{\rho} - I_{max}^{\delta'-\epsilon'}(X_{10}:Z)_{\rho} - I_{max}^{\delta'-\epsilon'}(X_{11}:ZX_{10}X_{20})_{\rho}$$
$$- I_{max}^{\delta'-\epsilon'}(X_{20}:ZX_{10})_{\rho} - I_{max}^{\delta'-\epsilon'}(X_{22}:ZX_{10}X_{11}X_{20})_{\rho} - 4\log\frac{3}{\epsilon'^3} + \log\delta' + 2\log\epsilon - 4 + \mathcal{O}(1) \quad (12)$$

$$2R_1 + R_2 \leq I_H^{\epsilon}(X_{11}:Y_1X_{10}X_{20})_{\rho} + I_H^{\epsilon}(X_{10}X_{22}:Y_2X_{20})_{\rho} + I_H^{\epsilon}(X_{11}X_{10}X_{20}:Y_2)_{\rho} - 2I_{max}^{\delta'-\epsilon'}(X_{10}:Z)_{\rho}$$
$$- 2I_{max}^{\delta'-\epsilon'}(X_{11}:ZX_{10}X_{20})_{\rho} - I_{max}^{\delta'-\epsilon'}(X_{20}:ZX_{10})_{\rho} - I_{max}^{\delta'-\epsilon'}(X_{22}:ZX_{10}X_{11}X_{20})_{\rho} - 6\log\frac{3}{\epsilon'^3}$$
$$+ \frac{3}{2}\log\delta' + 3\log\epsilon - 6 + \mathcal{O}(1) \quad (13)$$

$$R_1 + 2R_2 \leq I_H^{\epsilon}(X_{11}X_{20}:Y_1X_{10})_{\rho} + I_H^{\epsilon}(X_{22}:Y_2X_{10}X_{20})_{\rho} + I_H^{\epsilon}(X_{22}X_{20}X_{10}:Y_1)_{\rho} - I_{max}^{\delta'-\epsilon'}(X_{10}:Z)_{\rho}$$
$$- I_{max}^{\delta'-\epsilon'}(X_{11}:ZX_{10}X_{20})_{\rho} - 2I_{max}^{\delta'-\epsilon'}(X_{20}:ZX_{10})_{\rho} - 2I_{max}^{\delta'-\epsilon'}(X_{22}:ZX_{10}X_{11}X_{20})_{\rho} - 6\log\frac{3}{\epsilon'^3}$$
$$+ \frac{3}{2}\log\delta' + 3\log\epsilon - 6 + \mathcal{O}(1) \quad (14)$$

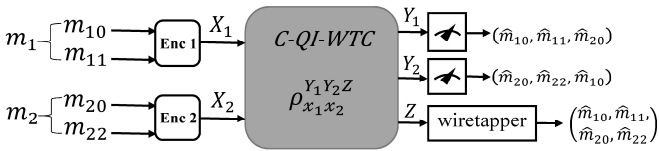

Figure 2. The structure of the C-QI-WTC under the Han-Kobayashi settings.

**Remark 1**: Note that, to take the intersection of the private regions for two 3-sender MACs raised in Theorem 1, we used the method of [26]. Another approach can be using *Fürier-Motzkin elimination* [Appendix D, 26] which gives achievable rate region similar to the Han-Kobayashi expression.

**Remark 2**: The Han-Kobayashi technique is based on rate splitting. It should be noted that the split messages are not independent of each other. Thus, obtaining secrecy against the eavesdropper by Wyner's randomizing technique becomes problematic in this setting. In other words, we cannot randomize over a block independently. For example, $m_1$ should be randomized using the product of two junk variables $(k_{10}, k_{11})$.

***Conjecture:*** *(An inner bound on the one-shot secrecy capacity region of the C-QI-WTC)* Consider the region:

$$\mathcal{R}(N) = \bigcup_{\pi}\{(R_1, R_2) \in R^2 | Eqns.(6) - (14) \; hold\}$$

*Proof:* In Appendix B.

*Sketch of proof:* We consider two sub C-QMA-WTCs. Therefore, from the perspective of the first receiver ($Y_1$), there are three messages $(m_{10}, m_{11}, m_{20}) \to (Y_1, Z)$, and for the second receiver, there are three messages $(m_{20}, m_{22}, m_{10}) \to (Y_2, Z)$. The paper [27] introduces the same setting, but it considers a randomized order such as $m_{10} \to m_{20} \to m_{11}$. For the first C-QMA-WTC, Alice should randomize over a total block of size $(k_{10}.k_{11})$. For the second C-QMA-WTC, Bob should randomize over a total block of size $(k_{20}.k_{22})$. Then, we can analyze both sub-channels.

***Remark 3***: The above conjecture holds *if and only if* condition (5) holds. Because taking the intersection of the private regions for two 3-sender C-QMACs is not enough to get a private region for the full C-QI-WTC.

To overcome the above problem, we should change the encoding process, which results in the following theorem.

***Theorem 2***: *(An inner bound on the one-shot secrecy capacity region of the C-QI-WTC) Consider the region:*

$$\mathcal{R}(N) = \bigcup_\pi \{(R_1, R_2) \in R^2 | Eqns.(15) - (28)\ hold\}$$

*Proof*: In Appendix C.

*Sketch of proof*: The overall sketch of the proof is the same as that for the above Conjecture with one difference: Suppose that both receivers want to decode non-interfering messages. Also, this setting is similar to Theorem 1. It can be helpful for the receivers to decode their messages, including the intended messages and interfering messages. In other words, $X_{10}$ and $X_{20}$ can be used as side information. Therefore, the first sub-channel can be modeled as $(X_{10}X_{11}X_2) \to (Y_1, Z)$. All steps, such as encoding and decoding, are the same as for the above Conjecture.

*Secrecy criterion*: The secrecy criterion for the channel can be defined as follows:

$$I(M_1, M_2: Z) \leq \vartheta \quad \text{For Theorem 1}$$

$$I(M_{10}, M_{11}, M_{20}, M_{22}: Z) \leq \vartheta \quad \text{For Conjecture and Theorem 2}$$

This means that the mutual information between the sent messages and the wiretapper should be bounded above by an arbitrarily small number.

## V. DISCUSSION AND FUTURE WORKS

In this paper, the problem of secure communication over a quantum interference channel has been studied. The main approach for decoding sent messages is simultaneous decoding (one-shot quantum joint typicality lemma) [19]. Also, we used the method of [27] to randomize Eve's knowledge and calculate leaked information. The mentioned Conjecture gives a one-shot achievable rate region for C-QI-WTC in the form of the Han-Kobayashi rate region. Still, it is not clear how we can conclude secrecy requirement for this channel from secrecy criterion of sub C-QMA-WTCs. However, Theorem 2 solves this problem using a new encoding.

## APPENDIX

***Appendix A:*** *(Proof of the Theorem 1)*

The channel in the Figure 1 can be split into two sub-QMA-WTCs with classical inputs. One from both users to $(Y_1, Z)$ and another from both users to $(Y_2, Z)$. At last, the overall achievable secrecy rate region can be calculated as:

$$\mathcal{R}_{C-QI-WTC} \leq \min\{\mathcal{R}_{C-QMA-WTC_1}, \mathcal{R}_{C-QMA-WTC_2}\}$$

Consider the first sub-channel. From Sen's jointly typical decoder [19] and [Lemma 3.2, 27], it is clear that:

$$R_1 \leq I_H^\epsilon(X_1: X_2Y_1|Q)_\rho - I_{max}^\eta(X_1: Z|Q)_\rho + \log \epsilon - 1 - \log \frac{3}{\epsilon'^3}$$
$$+ \frac{1}{4} \log \delta$$

$$R_2 \leq I_H^\epsilon(X_2: X_1Y_1|Q)_\rho - I_{max}^\eta(X_1: ZX_2|Q)_\rho + \log \epsilon - 1$$
$$- \log \frac{3}{\epsilon'^3} + \frac{1}{4} \log \delta$$

$$R_1 + R_2 \leq I_H^\epsilon(X_1X_2: Y_1|Q)_\rho - I_{max}^\eta(X_1: Z|Q)_\rho$$
$$- I_{max}^\eta(X_1: ZX_2|Q)_\rho + \log \epsilon - 1 - 2\log \frac{3}{\epsilon'^3}$$
$$+ \frac{1}{2} \log \delta + \mathcal{O}(1)$$

There are similar rates for the second sub-channel. Taking the intersection of the derived regions for the two sub-channels completes the proof.

*Secrecy criterion:* The secrecy constraint requires that Eve just could be able to decode a negligible information:

$$I_{max}^\eta(M_1M_2: Z)_\rho \leq \vartheta$$

It is obvious that [Lemma 3.2, 27] guarantees the secrecy criterion.

***Appendix B:*** *(Proof of the Conjecture)*

To bypass the problem raised in Remark 1 and recover the non-corner points in the secrecy rate region, we use rate splitting. We apply the following setting:

We consider two sub C-QMA-WTCs. Therefore, from the perspective of the first receiver $(Y_1)$, there are three messages $(m_{10}, m_{11}, m_{20}) \to (Y_1, Z)$, and for the second receiver, there are three messages $(m_{20}, m_{22}, m_{10}) \to (Y_2, Z)$. The paper [27] introduces the same setting, but it considers a randomized order such as $m_{10} \to m_{20} \to m_{11}$. This order has not impact on decoding the messages, but it is helpful to compute leaked information. Also, it should be considered that in the one-shot case, we do not use the successive decoder because the time-sharing strategy gives only finite achievable rate pair. Instead, we use the one-shot jointly typical decoder [19] for both sub-channels.

For the first C-QMA-WTC, Alice should randomize over total block of size $(k_{10}.k_{11})$. It refers to the fact that the split messages are dependent. There is a detailed discussion in [28].

For the C-QI-WTC, the controlling state is as follows:

$$R_1 \leq \min\{I_H^\epsilon(X_{10}X_{11}:Y_1X_2)_\rho, I_H^\epsilon(X_1:Y_2X_{20}X_{22})_\rho\} - I_{max}^{\delta'-\epsilon'}(X_{10}:Z)_\rho - I_{max}^{\delta'-\epsilon'}(X_{11}:ZX_{10}X_{20})_\rho - 2\log\frac{3}{\epsilon'^3}$$
$$+ \frac{1}{2}\log\delta' + \log\epsilon - 2 + \mathcal{O}(1) \tag{15}$$

$$R_1 \leq \{I_H^\epsilon(X_{11}:Y_1X_{10}X_2)_\rho + I_H^\epsilon(X_{10}:Y_1X_{11}X_2)_\rho, I_H^\epsilon(X_1X_{20}:Y_2X_{22})_\rho, I_H^\epsilon(X_1X_{22}:Y_2X_{20})_\rho\} - I_{max}^{\delta'-\epsilon'}(X_{10}:Z)_\rho$$
$$- I_{max}^{\delta'-\epsilon'}(X_{11}:ZX_{10}X_{20})_\rho - 2\log\frac{3}{\epsilon'^3} + \frac{1}{2}\log\delta' + \log\epsilon - 2 + \mathcal{O}(1) \tag{16-17}$$

$$R_2 \leq \min\{I_H^\epsilon(X_{20}X_{22}:Y_2X_1)_\rho, I_H^\epsilon(X_2:Y_1X_{10}X_{11})_\rho\} - I_H^\epsilon(X_{20}X_{22}:Y_2X_{10})_\rho - I_{max}^{\delta'-\epsilon'}(X_{20}:ZX_{10})_\rho$$
$$- I_{max}^{\delta'-\epsilon'}(X_{22}:ZX_{10}X_{11}X_{20})_\rho - 2\log\frac{3}{\epsilon'^3} + \frac{1}{2}\log\delta' + \log\epsilon - 2 + \mathcal{O}(1) \tag{18}$$

$$R_2 \leq \{I_H^\epsilon(X_{22}:Y_2X_{20}X_1)_\rho + I_H^\epsilon(X_{20}:Y_2X_{22}X_1)_\rho, I_H^\epsilon(X_2X_{10}:Y_1X_{11})_\rho, I_H^\epsilon(X_2X_{11}:Y_1X_{10})_\rho\} - I_{max}^{\delta'-\epsilon'}(X_{20}:ZX_{10})_\rho$$
$$- I_{max}^{\delta'-\epsilon'}(X_{22}:ZX_{10}X_{11}X_{20})_\rho - 2\log\frac{3}{\epsilon'^3} + \frac{1}{2}\log\delta' + 2\log\epsilon - 4 + \mathcal{O}(1) \tag{19-21}$$

$$R_1 + R_2 \leq \min\{I_H^\epsilon(X_{11}X_{10}X_2:Y_1)_\rho, I_H^\epsilon(X_{22}X_{20}X_1:Y_2)_\rho\} - I_{max}^{\delta'-\epsilon'}(X_{10}:Z)_\rho - I_{max}^{\delta'-\epsilon'}(X_{11}:ZX_{10}X_2)_\rho$$
$$- I_{max}^{\delta'-\epsilon'}(X_{20}:ZX_{10})_\rho - I_{max}^{\delta'-\epsilon'}(X_{22}:ZX_1X_{20})_\rho - 4\log\frac{3}{\epsilon'^3} + \log\delta' + 2\log\epsilon - 4 + \mathcal{O}(1) \tag{22}$$

$$R_1 + R_2 \leq \{I_H^\epsilon(X_{11}:Y_1X_{10}X_2)_\rho + I_H^\epsilon(X_{10}X_{20}:Y_1X_{11})_\rho, I_H^\epsilon(X_{22}:Y_2X_{20}X_1)_\rho + I_H^\epsilon(X_{20}X_{10}:Y_1X_{11})_\rho, I_H^\epsilon(X_{22}X_1:Y_2X_{20})_\rho$$
$$+ I_H^\epsilon(X_{20}:X_{22}X_1Y_2)_\rho, I_H^\epsilon(X_{11}X_2:Y_1X_{10})_\rho + I_H^\epsilon(X_{10}:X_{11}X_2Y_1)_\rho\} - I_{max}^{\delta'-\epsilon'}(X_{10}:Z)_\rho$$
$$- I_{max}^{\delta'-\epsilon'}(X_{11}:ZX_{20}X_{10})_\rho - I_{max}^{\delta'-\epsilon'}(X_{20}:ZX_{10})_\rho - I_{max}^{\delta'-\epsilon'}(X_{22}:ZX_{10}X_{11}X_{20})_\rho - 4\log\frac{3}{\epsilon'^3} + \log\delta'$$
$$+ 2\log\epsilon - 4 + \mathcal{O}(1) \tag{23-26}$$

$$2R_1 + R_2 \leq I_H^\epsilon(X_{20}X_1:Y_2X_{22})_\rho + I_H^\epsilon(X_1X_{22}:Y_2X_{20})_\rho - 2I_{max}^{\delta'-\epsilon'}(X_{10}:Z)_\rho - 2I_{max}^{\delta'-\epsilon'}(X_{11}:ZX_{10}X_{20})_\rho$$
$$- I_{max}^{\delta'-\epsilon'}(X_{20}:ZX_{10})_\rho - I_{max}^{\delta'-\epsilon'}(X_{22}:ZX_{10}X_{11}X_{20})_\rho - 6\log\frac{3}{\epsilon'^3} + \frac{3}{2}\log\delta' + 2\log\epsilon - 4 + \mathcal{O}(1) \tag{27}$$

$$R_1 + 2R_2 \leq I_H^\epsilon(X_{10}X_2:Y_1X_{11})_\rho + I_H^\epsilon(X_2X_{11}:Y_1X_{10})_\rho - I_{max}^{\delta'-\epsilon'}(X_{10}:Z)_\rho - I_{max}^{\delta'-\epsilon'}(X_{11}:ZX_{10}X_2)_\rho$$
$$- 2I_{max}^{\delta'-\epsilon'}(X_{20}:ZX_{10})_\rho - 2I_{max}^{\delta'-\epsilon'}(X_{22}:ZX_{10}X_{11}X_{20})_\rho - 6\log\frac{3}{\epsilon'^3} + \frac{3}{2}\log\delta' + +2\log\epsilon - 4$$
$$+ \mathcal{O}(1) \tag{28}$$

---

$$\rho^{X_{10}X_{11}X_{20}X_{22}Y_1Z}$$
$$:= \sum_{\substack{X_{10},X_{11}\in\mathcal{X}_1 \\ X_{20},X_{22}\in\mathcal{X}_2}} P_{X_{10}}(x_{10})P_{X_{20}}(x_{20})P_{X_{11}}(x_{11})P_{X_{22}}(x_{22})$$
$$|x_{10}\rangle\langle x_{10}|_{X_{10}} \otimes |x_{11}\rangle\langle x_{11}|_{X_{11}} \otimes |x_{20}\rangle\langle x_{20}|_{X_{20}}$$
$$\otimes |x_{22}\rangle\langle x_{22}|_{X_{22}} \otimes \rho_{Y_1Y_2Z}^{x_{01}x_{11}x_{20}x_{22}} \tag{29}$$

To simplify the analysis, we first remove the security constraint of the problem. From Sen's one-shot jointly typical decoder [19], we have the following region for the first C-QMAC:

$$R'_{10} \leq I_H^\epsilon(X_{10}:Y_1X_{11}X_{20})_\rho + \log\epsilon - 2$$
$$R'_{11} \leq I_H^\epsilon(X_{11}:Y_1X_{10}X_{20})_\rho + \log\epsilon - 2$$
$$R'_{20} \leq I_H^\epsilon(X_{20}:Y_1X_{10}X_{11})_\rho + \log\epsilon - 2$$
$$R'_{10} + R'_{11} \leq I_H^\epsilon(X_{10}X_{11}:Y_1X_{20})_\rho + \log\epsilon - 2$$
$$R'_{10} + R'_{20} \leq I_H^\epsilon(X_{10}X_{20}:Y_1X_{11})_\rho + \log\epsilon - 2$$
$$R'_{11} + R'_{20} \leq I_H^\epsilon(X_{11}X_{20}:Y_1X_{10})_\rho + \log\epsilon - 2$$
$$R'_{10} + R'_{11} + R'_{20} \leq I_H^\epsilon(X_{10}X_{11}X_{20}:Y_1)_\rho + \log\epsilon - 2$$

Also, for the second C-QMAC there are similar rates. It should be noted that $R_1 = R_{10} + R_{11}$ and $R_2 = R_{20} + R_{22}$. After eliminating redundant rates and using the *Fürier-Motzkin elimination,* we have:

$$\mathcal{R}_{C-QIC} = \bigcup_{\pi: P_{X_{10}}(x_{10})P_{X_{11}}(x_{11})P_{X_{20}}(x_{20})P_{X_{22}}(x_{22})}$$

$$R'_1 \leq I_H^\epsilon(X_{10}X_{11}:Y_1X_{20})_\rho + \log\epsilon - 2$$
$$R'_1 \leq I_H^\epsilon(X_{11}:Y_1X_{10}X_{20})_\rho + I_H^\epsilon(X_{10}:Y_2X_{20}X_{22})_\rho + 2\log\epsilon - 4$$
$$R'_2 \leq I_H^\epsilon(X_{20}X_{22}:Y_2X_{10})_\rho + \log\epsilon - 2$$
$$R'_2 \leq I_H^\epsilon(X_{20}:Y_1X_{10}X_{11})_\rho + I_H^\epsilon(X_{22}:Y_2X_{10}X_{20})_\rho + 2\log\epsilon - 4$$
$$R'_1 + R'_2 \leq I_H^\epsilon(X_{11}:Y_2X_{10}X_{20})_\rho + I_H^\epsilon(X_{10}X_{11}X_{20}:Y_2)_\rho + 2\log\epsilon - 4$$
$$R'_1 + R'_2 \leq I_H^\epsilon(X_{11}X_{20}:Y_1X_{10})_\rho + I_H^\epsilon(X_{22}X_{10}:Y_2X_{20})_\rho + 2\log\epsilon - 4$$

$$R_1 \leq I_H^\epsilon(X_{10}X_{11}:Y_1X_2)_\rho - I_{max}^{\delta'-\epsilon'}(X_{10}:Z)_\rho - I_{max}^{\delta'-\epsilon'}(X_{11}:ZX_{10}X_2)_\rho - 2\log\frac{3}{\epsilon'^3} + \frac{1}{2}\log\delta' + \log\epsilon - 2 + \mathcal{O}(1) \tag{30}$$

$$R_1 \leq I_H^\epsilon(X_{11}:Y_1X_{10}X_2)_\rho + I_H^\epsilon(X_{10}:Y_1X_{11}X_2)_\rho - I_{max}^{\delta'-\epsilon'}(X_{10}:Z)_\rho - I_{max}^{\delta'-\epsilon'}(X_{11}:ZX_{10}X_2)_\rho - 2\log\frac{3}{\epsilon'^3} + \frac{1}{2}\log\delta'$$
$$+ 2\log\epsilon - 4 + \mathcal{O}(1) \tag{31}$$

$$R_2 \leq I_H^\epsilon(X_2:Y_1X_{10}X_{11})_\rho - I_{max}^{\delta'-\epsilon'}(X_{20}:ZX_{10})_\rho - I_{max}^{\delta'-\epsilon'}(X_{22}:ZX_{10}X_{11}X_{20})_\rho - 2\log\frac{3}{\epsilon'^3} + \frac{1}{2}\log\delta' + \log\epsilon - 2$$
$$+ \mathcal{O}(1) \tag{32}$$

$$R_2 \leq I_H^\epsilon(X_2X_{10}:Y_1X_{11})_\rho - I_{max}^{\delta'-\epsilon'}(X_{20}:ZX_{10})_\rho - I_{max}^{\delta'-\epsilon'}(X_{22}:ZX_{10}X_{11}X_{20})_\rho - 2\log\frac{3}{\epsilon'^3} + \frac{1}{2}\log\delta' + \log\epsilon - 2$$
$$+ \mathcal{O}(1) \tag{33}$$

$$R_2 \leq I_H^\epsilon(X_2X_{11}:Y_1X_{10})_\rho - I_{max}^{\delta'-\epsilon'}(X_{20}:ZX_{10})_\rho - I_{max}^{\delta'-\epsilon'}(X_{22}:ZX_{10}X_{11}X_{20})_\rho - 2\log\frac{3}{\epsilon'^3} + \frac{1}{2}\log\delta' + \log\epsilon - 2$$
$$+ \mathcal{O}(1) \tag{34}$$

$$R_1 + R_2 \leq I_H^\epsilon(X_{11}:Y_1X_{10}X_2)_\rho + I_H^\epsilon(X_{10}X_{20}:Y_1X_{11})_\rho - I_{max}^{\delta'-\epsilon'}(X_{10}:Z)_\rho - I_{max}^{\delta'-\epsilon'}(X_{11}:ZX_{10}X_2)_\rho$$
$$- I_{max}^{\delta'-\epsilon'}(X_{20}:ZX_{10})_\rho - I_{max}^{\delta'-\epsilon'}(X_{22}:ZX_{10}X_{11}X_{20})_\rho - 4\log\frac{3}{\epsilon'^3} + \log\delta' + 2\log\epsilon - 4 + \mathcal{O}(1) \tag{35}$$

$$R_1 + R_2 \leq I_H^\epsilon(X_{11}X_2:Y_1X_{10})_\rho + I_H^\epsilon(X_{10}:X_{11}X_2Y_1)_\rho - I_{max}^{\delta'-\epsilon'}(X_{10}:Z)_\rho - I_{max}^{\delta'-\epsilon'}(X_{11}:ZX_{10}X_2)_\rho - I_{max}^{\delta'-\epsilon'}(X_{20}:ZX_{10})_\rho$$
$$- I_{max}^{\delta'-\epsilon'}(X_{22}:ZX_{10}X_{11}X_{20})_\rho - 4\log\frac{3}{\epsilon'^3} + \log\delta' + 2\log\epsilon - 4 + \mathcal{O}(1) \tag{36}$$

$$R_1 + R_2 \leq I_H^\epsilon(X_{11}X_{10}X_2:Y_1)_\rho - I_{max}^{\delta'-\epsilon'}(X_{10}:Z)_\rho - I_{max}^{\delta'-\epsilon'}(X_{11}:ZX_{10}X_2)_\rho - I_{max}^{\delta'-\epsilon'}(X_{20}:ZX_{10})_\rho$$
$$- I_{max}^{\delta'-\epsilon'}(X_{22}:ZX_{10}X_{11}X_{20})_\rho - 4\log\frac{3}{\epsilon'^3} + \log\delta' + \log\epsilon - 2 + \mathcal{O}(1) \tag{37}$$

$$R_1 + 2R_2 \leq I_H^\epsilon(X_{10}X_2:Y_1X_{11})_\rho + I_H^\epsilon(X_2X_{11}:Y_1X_{10})_\rho - I_{max}^{\delta'-\epsilon'}(X_{10}:Z)_\rho - I_{max}^{\delta'-\epsilon'}(X_{11}:ZX_{10}X_2)_\rho$$
$$- 2I_{max}^{\delta'-\epsilon'}(X_{20}:ZX_{10})_\rho - 2I_{max}^{\delta'-\epsilon'}(X_{22}:ZX_{10}X_{11}X_{20})_\rho - 6\log\frac{3}{\epsilon'^3} + \frac{3}{2}\log\delta' + +2\log\epsilon - 4$$
$$+ \mathcal{O}(1) \tag{38}$$

---

$$R_1' + R_2' \leq I_H^\epsilon(X_{11}:Y_1X_{20}X_{10})_\rho + I_H^\epsilon(X_{22}X_{20}X_{10}:Y_2)_\rho$$
$$+ 2\log\epsilon - 4$$

$$2R_1' + R_2' \leq I_H^\epsilon(X_{11}:Y_1X_{10}X_{20})_\rho + I_H^\epsilon(X_{10}X_{22}:Y_2X_{20})_\rho$$
$$+ I_H^\epsilon(X_{11}X_{10}X_{20}:Y_2)_\rho + 3\log\epsilon - 6$$

$$R_1' + 2R_2' \leq I_H^\epsilon(X_{11}X_{20}:Y_1X_{10})_\rho + I_H^\epsilon(X_{22}:Y_2X_{10}X_{20})_\rho$$
$$+ I_H^\epsilon(X_{22}X_{20}X_{10}:Y_1)_\rho + 3\log\epsilon - 6$$

This region is called the *quantum one-shot Han-Kobayashi rate region for C-QIC,* which is calculated in a special case by Sen [30]. He considers the case of an interference channel with independent prior entanglement between sender 1 and its intended receiver and between sender 2 and its intended receiver. It should be noted that the quantum *Han-Kobayashi rate region* for C-QIC in the i.i.d. case is conjectured in [14].

Note that, all of the above rates correspond to the C-QMACs without secrecy constraints. Now we want to consider the secrecy requirements of the problem.

For a C-QMA-WTC, we need a *smooth* version of the tripartite convex split lemma [20]. This runs into the smoothing bottleneck of quantum information theory. In [27], the authors suggested a novel lemma that gives the size of the randomized block in terms of smooth max mutual information.

**Lemma 1**: *Given the control state in (29) and decoding order such as $m_{10} \to m_{20} \to m_{11} \to m_{22}$, $\delta' > 0$ and $0 < \epsilon' < \delta'$, let $\{x_{10}^{(1)}, \dots, x_{10}^{(K_{10})}\}$, $\{x_{11}^{(1)}, \dots, x_{11}^{(K_{11})}\}$ and $\{x_2^{(1)}, \dots, x_2^{(K_2)}\}$ be i.i.d. samples from the distributions $P_{X_{10}}$, $P_{X_{11}}$ and $P_{X_2}$. Then, if*

$$\log|\mathcal{K}_{10}| \geq I_{max}^{\delta'-\epsilon'}(X_{10}:Z)_\rho + \log\frac{3}{\epsilon'^3} - \frac{1}{4}\log\delta'$$

$$\log|\mathcal{K}_{20}| \geq I_{max}^{\delta'-\epsilon'}(X_{20}:ZX_{10})_\rho + \log\frac{3}{\epsilon'^3} - \frac{1}{4}\log\delta' + \mathcal{O}(1)$$

$$\log|\mathcal{K}_{11}| \geq I_{max}^{\delta'-\epsilon'}(X_{11}:ZX_{10}X_{20})_\rho + \log\frac{3}{\epsilon'^3} - \frac{1}{4}\log\delta'$$
$$+ \mathcal{O}(1)$$

$$\log|\mathcal{K}_{22}| \geq I_{max}^{\delta'-\epsilon'}(X_{22}:ZX_{10}X_{11}X_{20})_\rho + \log\frac{3}{\epsilon'^3} - \frac{1}{4}\log\delta'$$
$$+ \mathcal{O}(1)$$

the following holds,

$$\mathbb{E}_{\substack{X_{10}\sim P_{X_{10}}\\X_{11}\sim P_{X_{11}}\\X_{20}\sim P_{X_{20}}\\X_{22}\sim P_{X_{22}}}}\left\|\frac{1}{|\mathcal{K}_1||\mathcal{K}_2|}\sum_{k=1}^{|\mathcal{K}_{22}|}\sum_{l=1}^{|\mathcal{K}_{11}|}\sum_{j=1}^{|\mathcal{K}_{20}|}\sum_{i=1}^{|\mathcal{K}_{10}|}\rho^Z_{x_{10}^i x_{20}^j x_{11}^l x_{22}^k} - \rho^Z\right\|_1$$
$$\leq 60\delta'^{\frac{1}{8}}$$

*Proof*: The proof is similar to the two-user case explained in [27].

As mentioned before, let $k_1 = k_{10}.k_{11}$ and $k_2 = k_{20}.k_{22}$. Note that, $R_1 = R_1' - \log k_1$, $R_2 = R_2' - \log k_2$. Using the above lemma completes the proof.

***Appendix C***: *(Proof of the Theorem 2)* As mentioned in Appendix A, the secrecy constraint requires that Eve just could be able to decode a negligible information:

$$I_{max}^\eta(m_{10}m_{11}m_{20}m_{22}:Z)_\rho \leq \vartheta \quad (39)$$

*Encoding*: Suppose that both receivers want to decode non-interfering messages. This setting is similar to Theorem 1. It can be helpful for the receivers to decode their messages, including the intended messages and interfering messages. In other words, $X_{10}$ and $X_{20}$ can be used as side information. Therefore, the first sub-channel can be modeled as $(X_{10}X_{11}X_2) \to (Y_1, Z)$.

Consider the first C-QMA-WTC $(X_{10}X_{11}X_2) \to (Y_1, Z)$. From [27], we know that an achievable rate region can be calculated as stated in (30)-(38).

For the second C-QMA-WTC $(X_1X_{20}X_{22}) \to (Y_2, Z)$, there are similar achievable rates. Taking the intersection of the secrecy regions for both sub-channels can be calculated as stated in (15)-(28). Against Conjecture, Lemma 1 guarantees that the secrecy constraint for this problem (39) holds. This completes the proof.